\documentclass[twocolumn,10pt]{article}

\usepackage[english]{babel}
\usepackage{hyperref}
\usepackage{doi}
\usepackage[switch]{lineno}
\usepackage{soul}

\usepackage[a4paper,margin=1.6cm]{geometry}
\usepackage{times}            
\usepackage{amsmath,amssymb}
\usepackage{graphicx}
\usepackage{physics}
\usepackage{siunitx}
\usepackage{authblk}          
\usepackage{titlesec}         
\usepackage{cite}
\usepackage{setspace}

\usepackage{tikz}
\usetikzlibrary{arrows.meta,positioning,shapes.geometric,external}
\usetikzlibrary{arrows.meta, positioning, calc, decorations.pathreplacing}

\usepackage{multirow} 

\titleformat{\section}{\bfseries\large}{\thesection}{0.5em}{} 
\titleformat{\subsection}{\bfseries}{\thesubsection}{0.5em}{} 

\title{\bfseries An Introduction to the Foundations and Interpretations of\\ Quantum Mechanics}

\author[1]{Theodore McKeever}
\author[1]{Ahsan Nazir}

\affil[1]{\small Department of Physics and Astronomy, The University of Manchester, Oxford Road, Manchester, M13 9PL, United Kingdom}

\date{} 

\begin{document}
\twocolumn[
\maketitle

\begin{abstract}
\noindent
This article surveys a selection of key conceptual and interpretational developments in quantum mechanics, tracing the theory from its foundational postulates to contemporary discussions of measurement, nonlocality, and the emergence of classicality. Beginning with the structure of Hilbert space and the postulates governing state evolution and measurement,  the epistemic stance of the Copenhagen interpretation and its modern reformulations are examined. The Einstein--Podolsky--Rosen argument, Bell’s theorem, and Hardy’s paradox are then discussed as probes of locality and realism, alongside the deterministic but explicitly nonlocal de Broglie--Bohm theory. The measurement problem and the implications of contextuality are analyzed in relation to objective collapse models, which introduce new physical dynamics to account for definite outcomes. Finally, the role of decoherence in the suppression of interference and the emergence of classical behavior is explored, together with the interpretational frameworks of many-worlds and consistent histories. This material aims to provide a coherent introductory overview of how several of the most prominent interpretations address the central concern of what quantum mechanics tells us about the nature of physical reality.
\end{abstract}
\vspace{1.5em}
]



\section{Introduction}\label{sec:intro}

Quantum mechanics has long been regarded not only as one of the most successful theories in physics, but also as one of the most philosophically perplexing. Over the last century, physicists and philosophers have proposed a wide range of interpretations in an effort to make sense of its mathematical formalism and empirical success.

This article surveys key developments in the closely related fields of quantum foundations—concerned with the axioms, structure, constraints, and empirical testability of quantum theory—and quantum philosophy, which examines the associated metaphysical and epistemological implications of available interpretations. Although these topics have been extensively reviewed in a variety of styles \cite{Bell_Aspect_2004,BungeSurvey1956,rovelli2025fourways,french2022interpretation,collins2007manyinterpretations,norsen2017foundations,Adlam_2021,encyclopedia2020071,interpPickl,sabathiel2014interpretation,Kok2023,davies_brown_1986,leifer2019foundationsnotes,lazarou2009interpretationquantumtheory,Obsorn1997quantum,tumulka2017quantmech},
the aim of this paper is to fill a more focused niche: to provide a contemporary, concise, accessible, and physically-grounded introduction to the central interpretational issues and results that shape the field. It is intended as a conceptual guide and entry point for readers with a background in quantum physics who seek familiarity with the landscape of its foundations. The material covered here, and the references provided, are non-exhaustive, serving more as a springboard towards further study rather than a complete record.

\begin{figure*}
\centering
\includegraphics[clip, trim=0cm 0.9cm 0.5cm 1.2cm,width=\textwidth]{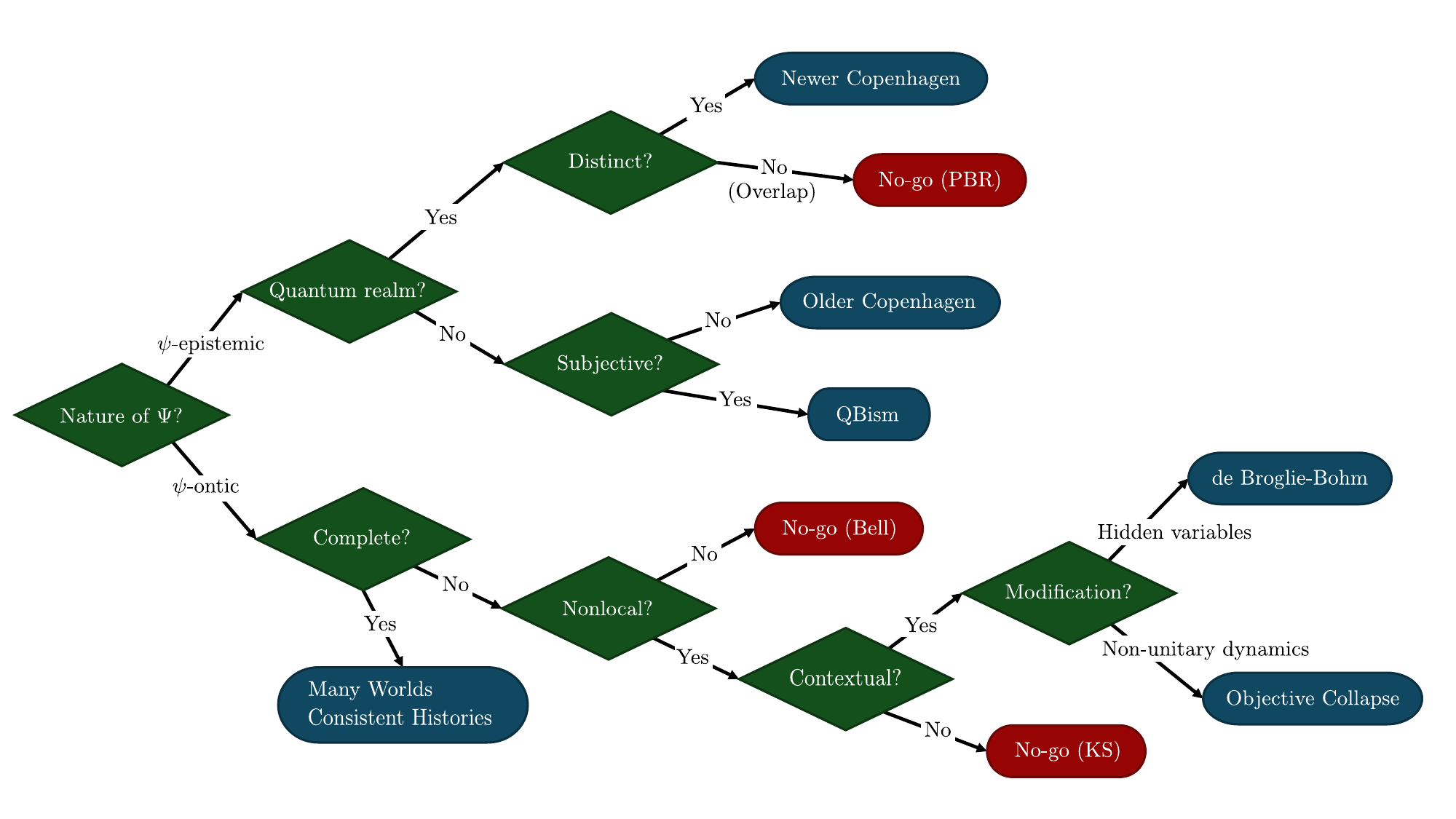}
\caption{A decision tree broadly summarizing the quantum-mechanical interpretations and no-go results discussed in the main text, and mirroring its structure. Green diamonds denote conceptual questions, red boxes theory classes excluded by no-go theorems, and blue boxes a non-exhaustive selection of interpretations. Here, ``Quantum realm?'' asks whether the quantum formalism refers to an underlying ontological reality; ``Distinct?'' whether distinct quantum states may correspond to the same underlying physical state; ``Subjective?'' whether quantum probabilities are agent-dependent beliefs; ``Complete?'' whether the standard postulates of quantum mechanics are taken as complete; ``Nonlocal?'' and ``Contextual?'' refer respectively to Bell nonlocality and Kochen–Specker contextuality; and ``Modification?'' asks how standard quantum mechanics is altered.}
\label{fig:decision-tree}
\end{figure*}

The guiding questions that motivate this discussion are profound. Is the quantum state an element of physical reality, or merely a compact way of expressing what we know about a system? Is the theory as it stands complete, or must it be supplemented with additional (hidden) variables? Do quantum systems have definite properties before we measure them? And if not, how do we draw the line between the classical world of experience and the underlying quantum description of nature? Why does quantum theory describe two very different types of time evolution—smooth and reversible in ordinary dynamics, but discontinuous and irreversible during measurement? Finally, what does it actually mean to say that a measurement has taken place? In exploring these questions, we examine several of the most influential interpretations of quantum mechanics, highlighting their conceptual commitments, strengths, and limitations, and using key theorems and no-go results as guideposts in navigating the interpretational landscape. {Figure~\ref{fig:decision-tree} offers a schematic overview of the structure of our analysis.}

\section{The Postulates of Quantum Mechanics and the Copenhagen interpretation}
\subsection{The Postulates of Quantum Mechanics}\label{sec:postulates}

Quantum mechanics is built upon a set of (five) postulates \cite{Dirac1930,Pade2018}. These are not axioms in the strict mathematical sense of being logically minimal, but rather rules that tell us how to use the theory. Taken together, they provide a complete framework for predicting and analyzing experiments.

\subsection*{Postulate 1: States and State Space}

A physical system is described by a mathematical object called a state vector, $\ket{\psi}$, which is an element of a Hilbert space $\mathcal H$.\footnote{Strictly speaking, states are \emph{rays} rather than vectors, since an overall (global) phase $e^{i\varphi}$ has no physically observable consequences, and $e^{i\varphi}|\psi\rangle$ represents the same state as $|\psi\rangle$.} The state vector must be normalized such that $\langle \psi|\psi \rangle = 1$. Hilbert space is a vector space, which ensures that superpositions of states are well defined and that interference phenomena are possible. This mathematical structure guarantees that probabilities add up correctly and that different states can combine in physically meaningful ways. An interpretational question arises immediately: does the state vector correspond to something physically real about the system itself (an ontic view), or is it simply an expression of our knowledge or information about the system (an epistemic view)? 

{If a physical system is composed of two or more subsystems, with state spaces $\mathcal H_1$ and $\mathcal H_2$, then the state space of the composite system is the tensor product space $\mathcal H_1\otimes\mathcal H_2$. States of the joint system are therefore vectors in this larger Hilbert space. Some of these states can be written as product states, i.e., in the form $|\psi_1\rangle\otimes|\psi_2\rangle$, corresponding to independently specified subsystem states, but in general the joint state need not factorize. Non-factorizable states are called entangled states and play a central role in quantum phenomena.}

{Entanglement implies that the composite system may possess well-defined properties even when no definite pure state can be assigned to each subsystem individually.\footnote{In such cases, even when the total state is pure, the state of an individual subsystem is generally described by a mixed reduced state obtained by tracing over the other degrees of freedom (see density operators later in Section~\ref{subsec:densityOperators}).} This structural feature of the tensor-product state space underlies many of the non-classical correlations discussed later, including those appearing in the Einstein–Podolsky–Rosen argument \cite{EPR1935} and Bell-type experiments \cite{Bell1964}.} 

\subsection*{Postulate 2: Time Evolution}

The second postulate describes how states change over time. An isolated system evolves according to the Schrödinger equation \cite{TDSE1926,Griffiths_Schroeter_2018}, 
\begin{align}\label{eq:TDSE}
i \hbar \frac{d}{dt}\ket{\psi(t)} = H \ket{\psi(t)} ,
\end{align}
in which $H$ is the Hamiltonian, the Hermitian operator representing the energy of the system, and governs the time development of the state. This time evolution is unitary, meaning it preserves probabilities and is reversible. 

\subsection*{Postulate 3: Physical Observables}

The third postulate connects the abstract state vector to measurable physical quantities. Every observable, such as position, momentum, or spin, is represented by a Hermitian operator. Hermitian operators are significant because their eigenvalues are guaranteed to be real numbers, which can be associated with the possible outcomes of experiments. 
It is worth emphasizing that “observable” in this context does not imply the need for a conscious observer; it refers simply to any measurable physical quantity.

\subsection*{Postulate 4: Measurement Outcomes}

The fourth postulate explains how measurement outcomes are distributed. If we measure an observable $A$, the only possible results are the eigenvalues of $A$. The probability of obtaining a particular eigenvalue $a_j$ when the system is in state $\ket{\psi}$ is given by the Born rule \cite{Born1926QuantenmechanikDS,Landsman2009}:  
\begin{align}\label{eq:BornRule}
P(a_j) = \left| \langle a_j | \psi \rangle \right|^2
\end{align}
where $\ket{a_j}$ is the corresponding eigenstate of $A$. For Hilbert spaces of dimension greater than two, Gleason's theorem \cite{Gleason1957} shows that any consistent assignment of probabilities to quantum measurement outcomes must be representable in the Born-rule form.

A vivid example of this principle is the Stern--Gerlach experiment \cite{SternGerlach1922}, in which a beam of silver atoms passing through an inhomogeneous magnetic field splits into discrete components corresponding to the eigenvalues of a specified spin operator. Individual detection events are unpredictable, even for identically prepared states, yet repeated trials produce stable statistical frequencies. The experiment thus exemplifies both the quantization of observables and the irreducibly probabilistic character of quantum measurement.
This probabilistic structure has fueled much of the debate about whether quantum mechanics is complete. For example, do probabilities merely reflect our ignorance of hidden variables, or are they fundamental features of nature?

\subsection*{Postulate 5: Projection (Collapse)}

Finally, the fifth postulate describes what happens after a measurement. If a measurement of an observable $A$ yields the result $a_j$, then immediately afterwards the system is in the corresponding eigenstate  $\ket{a_j}$. This projection process is non--unitary and irreversible. It also raises the ``measurement problem'' (discussed in Section~\ref{sec:MeasurementProblem}): why should the system, together with the measuring apparatus (which are both physical systems), not continue to evolve according to the Schrödinger equation? Nevertheless, experiments always yield definite outcomes, so some version of the projection postulate seems unavoidable.   

\subsection*{Summary of Postulates}
{Taken together, these postulates define the operational framework of quantum mechanics: physical systems are represented by normalized state vectors in Hilbert space, composite systems by tensor-product spaces that admit entangled states, observables by Hermitian operators, and measurement outcomes by the Born rule together with state-update upon measurement. This structure has proven extraordinarily successful in practice; indeed, the pragmatic Copenhagen tradition emphasizes its empirical adequacy, taking the postulates largely at face value. At the same time, the postulates raise deep conceptual questions about what the quantum formalism is telling us about the nature of reality: for example, what does the state vector mean, why is a special status assigned to measurement, and should a fundamental physical theory tell us something more than just rules for predicting experimental outcomes? 
}


\subsection{The Copenhagen Interpretation}

To begin to address these puzzles, we turn first to the Copenhagen interpretation, or rather interpretations, since the term actually refers to a family of related but not always consistent viewpoints. Despite their differences, these interpretations share a number of common core principles. They deny the existence of hidden variables and regard the state vector as the most complete description possible of an individual system. They hold that measurement outcomes are objectively random, not simply the result of ignorance. Accordingly, they also insist that quantum mechanics applies to single systems rather than statistical ensembles, and they deny that unmeasured systems can be assigned definite properties.

\section{The PBR Argument and the Nature of Quantum States}\label{sec:PBR}

One of the central debates concerns the nature of the quantum state itself, and the question of whether it is \emph{$\psi$-epistemic} or \emph{$\psi$-ontic}.
In a $\psi$-ontic conception, distinct pure quantum states {represent} 
distinct physical states of the world, so that the quantum state uniquely characterizes the system’s physical situation, even if $\psi$ is not interpreted as a literal physical object. By contrast, a $\psi$-epistemic conception holds that the quantum state represents information, beliefs, or knowledge about an underlying physical reality, allowing in principle for distinct quantum states to correspond to overlapping descriptions of the same physical state. Whether the quantum state should be regarded as a direct representation of underlying reality or merely as a predictive tool is a further interpretational question. Notably, if a $\psi$-epistemic model holds that $\psi$ refers to an underlying reality and satisfies the condition that each complete physical state is compatible with only one pure quantum state, then it becomes empirically indistinguishable from a $\psi$-ontic view \cite{HarriganSpekkens2010}, differing only in metaphysical interpretation rather than physical content.



\subsection{The PBR Theorem}

\begin{figure}
    \centering
    \includegraphics[clip, trim=5.5cm 8.3cm 5cm 8.8cm,width=0.9\textwidth]{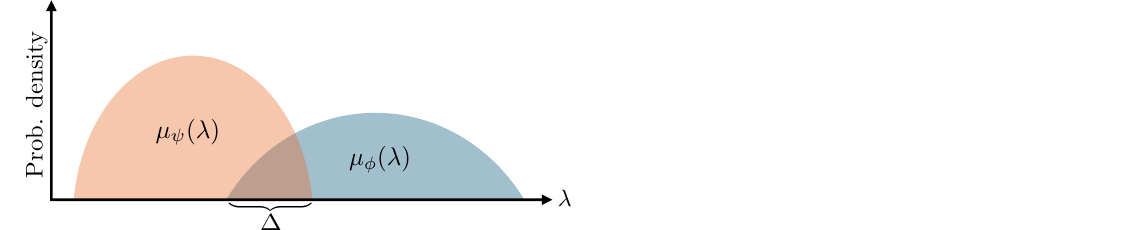}
    \caption{\label{fig:PBR}An illustration of $\psi$-overlap theories that PBR theorem aims to falsify (in the Popperian sense \cite{popper1959logic}), with supports $\mu_\psi (\lambda)$ and $\mu_\phi (\lambda)$ that both extend into the region $\Delta$.}
\end{figure}

\textit{Prima facie}, epistemic views are free to suggest that the state vector does not correspond directly to physical reality, but instead represents incomplete information about an underlying physical state. The Pusey--Barrett--Rudolph (PBR) theorem \cite{PBR2012} shows that a broad class of such epistemic hidden-variable models is incompatible with quantum mechanical predictions, provided one accepts a natural independence assumption about separately prepared systems.

To begin, one assumes that a system has an underlying physical (ontic) state $\lambda \in \Lambda$, where $\Lambda$ denotes the space of all possible complete physical states of the system, and that a preparation of a quantum state $\ket{\psi}$ corresponds to a probability distribution $\mu_\psi(\lambda)$ over $\Lambda$. Measurement outcomes are determined probabilistically by response functions $\xi(k|M,\lambda)$, where $k$ labels the outcome of a measurement $M$. The model reproduces quantum theory if the conditional probability
\[
P(k|M,\psi)=\int_\Lambda d\lambda\; \mu_\psi(\lambda)\,\xi(k|M,\lambda)
\]
agrees with the Born rule for all $\ket{\psi}$ and $M$.

In this setting, we call a model \emph{$\psi$-overlap} if distinct quantum states can correspond to overlapping ontic supports: there exist $\ket{\psi}\neq\ket{\phi}$ and a region $\Delta\subset\Lambda$
\[
\mu_\psi(\lambda)>0 \ \  \text{and}\ \ \mu_\phi(\lambda)>0 \quad \text{for}\ \lambda\in\Delta,
\]
so that the same underlying reality $\lambda$ could arise from more than one preparation, as depicted in Figure~\ref{fig:PBR}. This captures a standard sense in which the quantum state is epistemic. Not every $\psi$-epistemic stance, however, takes this form—for example, one may posit non-overlapping supports while still treating $\psi$ as informational, or reject the ontic-state framework entirely. The PBR theorem targets precisely the $\psi$-overlap class.


A key additional assumption used by PBR is \emph{preparation independence}: independently prepared systems possess independently distributed ontic states. PBR then show that if two distinct states $\ket{\psi}$ and $\ket{\phi}$ have overlapping ontic support, preparation independence implies that with nonzero probability two independently prepared systems will both have ontic states lying in the overlap region.

The argument proceeds, in the simplest illustration, by taking $\ket{\psi_0}=\ket{0}$ and $\ket{\psi_1}=\ket{+}=(\ket{0}+\ket{1})/\sqrt{2}$, and supposing their ontic distributions overlap with probability $q>0$. Preparing two systems independently, the probability that \emph{both} $\lambda$ lie in the overlap is at least $q^2$. On those occasions, the $\psi$-overlap view predicts an underlying 
physical state that is compatible with \emph{any} of the four product preparations
\[
\ket{\psi_{xy}} = \ket{\psi_x}\otimes\ket{\psi_y},\qquad x,y\in\{0,1\}.
\]

Crucially, the next step of PBR was to take a joint measurement on the basis of entangled pairs,\footnote{
The two-qubit entangled basis can be represented by: \mbox{$\ket{\xi_0}=\frac{1}{\sqrt{2}}\big(\!\ket{0}\!\otimes\!\ket{1}+\ket{1}\!\otimes\!\ket{0}\!\big)$}, \mbox{$\ket{\xi_1}=\frac{1}{\sqrt{2}}\big(\!\ket{0}\!\otimes\!\ket{-}+\ket{1}\!\otimes\!\ket{+}\!\big)$}, 
\mbox{$\ket{\xi_2}=\frac{1}{\sqrt{2}}\big(\!\ket{+}\!\otimes\!\ket{1}+\ket{-}\!\otimes\!\ket{0}\!\big)$},  
\mbox{$\ket{\xi_3}=\frac{1}{\sqrt{2}}\big(\!\ket{+}\!\otimes\!\ket{-}+\ket{-}\!\otimes\!\ket{+}\!\big)$}.
} $\{ \ket{\xi_z} \mid z=0,1,2,3 \}$. {By examining the inner products $\braket{\psi_{xy}}{\xi_z}$,} each outcome $\ket{\xi_z}$ is found to be orthogonal to, and thus excludes the measurement result of, a different $\ket{\psi_{xy}}$. Therefore, no such outcome exists that is compatible with all $\ket{\psi_{xy}}$. However, with probability $q^2$, the $\psi$-overlap view predicts a state that is precisely compatible with all $\ket{\psi_{xy}}$---an impossibility according to quantum theory. Thus, it is forbidden for the distributions of $\ket{0}$ and $\ket{+}$ to overlap. PBR subsequently extend this finding to arbitrary states $\ket{\psi_0}$ and $\ket{\psi_1}$ \cite{PBR2012}.

The PBR theorem therefore rules out $\psi$-overlap models, that is, $\psi$-epistemic models in which the quantum state represents incomplete information about an underlying physical reality shared across different preparations \cite{Leifer2014}. It does not, however, refute $\psi$-epistemic interpretations that treat the quantum state as complete, nor those that reject preparation independence or the appeal to a quantum reality altogether. Nevertheless, PBR provides a powerful no-go result: within a large and well-motivated class of hidden-variable theories, the quantum state cannot be merely a state of knowledge.

\subsection{Older Copenhagen}

In the older Copenhagen interpretation, shaped by Niels Bohr and Werner Heisenberg \cite{BOHR1928,Heisenberg1958}, the state vector is nothing more than a tool for describing phenomena, with no reference to an underlying reality\cite{Petersen01091963}:\footnote{This quote is often attributed to Bohr, though possibly mistakenly\cite{mermin2004whats}.}
\begin{quote}
There is no quantum world. There is only an abstract physical description. It is wrong to think that the task of physics is to find out how nature is. Physics concerns what we can say about nature.
\end{quote}

In this view, every measurement involves a division into two parts: the quantum system being observed, and the classical measuring apparatus that records the outcome. The apparatus, by necessity, is described in classical terms. Heisenberg formalized this division by introducing the notion of a ``cut'' between the quantum domain and the classical domain \cite{Heisenberg1930}. The location of this cut is not fixed, and can be shifted depending on the problem at hand, but it has no precise quantitative definition.

Bohr also introduced the principle of \emph{complementarity} \cite{BOHR1928}, which asserts that certain mutually exclusive descriptions are nevertheless jointly necessary for a full account of quantum phenomena, despite the absence of a universally agreed formal definition \cite{Complementarity1949}.
The wave and particle descriptions of light and matter are the most famous example. One cannot observe both simultaneously in a single experimental setup, yet both descriptions are required to capture the full range of phenomena. Complementarity is closely tied to Heisenberg’s uncertainty principle \cite{Heisenberg1927}, which shows that conjugate variables such as position and momentum cannot both be defined with arbitrary precision. Importantly, complementarity should not be understood simply as a limitation of measurement. It reflects a deeper structural feature of quantum theory: different experimental arrangements reveal mutually exclusive aspects of a system that cannot be jointly realized within a single experimental context.


The \emph{correspondence principle} \cite{Bohr1920} is a further cornerstone of the older Copenhagen interpretation, though it applies equally to any interpretation of quantum mechanics. It requires that, in the limit of large systems or large quantum numbers, quantum predictions approximate those of classical physics. Quantum mechanics does not reduce to classical mechanics in any exact sense, but it must reproduce classical behavior on average in the appropriate regime. This principle underlies the expectation that macroscopic systems behave classically, even though modern efforts to construct large-scale quantum technologies continue to test its limits.

\subsection{Newer Copenhagen}

The newer or standard Copenhagen interpretation is less radical than Bohr's original position. It accepts the real existence of atoms and electrons, and takes the postulates of quantum mechanics largely at face value. The quantum state remains epistemic in the sense that it represents the most complete description available for an individual system within the theory, but it is not treated as a direct representation of an underlying microscopic reality. Instead, the interpretation adopts a deliberately pragmatic stance, focusing on the operational use of the formalism. There is no sharp quantum-classical divide, but the collapse postulate is nevertheless invoked without explanation. This leaves us with the measurement problem, which we will return to in later sections. In practice, this interpretation aligns closely with the pragmatic stance of many physicists.

\subsection{Quantum Bayesianism (QBism)}

A more recent development is the approach known as Quantum Bayesianism, or QBism \cite{FuchsQBismIntro2014,mermin2012fixing}. This interpretation offers a modern refinement of the Copenhagen view by adopting a specifically Bayesian conception of probability. In the frequentist tradition, probabilities are defined by repeating an experiment many times and calculating the relative frequencies of outcomes. Bayesian probability, in contrast, is subjective: it represents the degree of belief that an agent assigns to a particular event. From this perspective, the quantum state is nothing more than a personal belief about the outcomes of future experiences. In QBism, whilst these beliefs are subjective, they are constrained by objective normative rules, including the Born rule \cite{FuchsQBismCoh2013}. 
Measurement is understood as an action performed by an agent, and the outcome is the agent’s new experience.

QBism thus takes an explicitly epistemic and agent-centered stance. Proponents argue that it provides a consistent way of understanding quantum mechanics without appealing to hidden variables or metaphysical assumptions. Critics, however, regard it as anti-realist, since it appears to deny that physics is about an external reality at all \cite{TIMPSON2008579}. Some proponents prefer to describe it as a form of participatory realism: the world is real, but quantum mechanics is not a mirror of reality; rather, it serves as a guide for each agent’s expectations concerning their interactions with the world.



\section{The EPR Argument and Nonlocality}\label{sec:EPR}

Having introduced the postulates of quantum mechanics and the interpretational perspectives that grew from the Copenhagen tradition, we now turn to {$\psi$-ontic models and} the question of whether the theory offers a complete description of physical reality. The apparent indeterminism of measurement and the probabilistic nature of the quantum state led many, including Einstein, to suspect that something essential was missing. {Moreover, the existence of entangled states leads to correlations between spacelike-separated systems that cannot be understood within classical probabilistic frameworks.} Section~\ref{sec:EPR} explores these questions through a series of arguments and theoretical developments that reveal deep tensions between locality, realism, and the predictions of quantum mechanics.

\subsection{The EPR Argument}

In 1935, Einstein, Podolsky, and Rosen (EPR) \cite{EPR1935} proposed a thought experiment intended to show that the quantum-mechanical description of nature is incomplete \cite{Wiseman_2014,norsen2017foundations,maudlin2011quantum}. Their reasoning began with the assumption that a \emph{complete} physical theory should account for all elements of physical reality. 
They then propose a sufficient condition for something to count as an element of reality: 
\begin{enumerate}
    \item[]\textbf{Reality:} If, without disturbing the system, we can predict with certainty the value of a quantity, then there exists an element of physical reality corresponding to that quantity.
\end{enumerate}
In a sense, if the system ``owns'' a property $X$ independent of measurement, this qualifies $X$ an element of physical reality.

The EPR argument can be framed as a trilemma between this \textbf{reality} condition and two other appealing propositions:
\begin{enumerate}
    \setcounter{enumi}{1}
    \item[] \textbf{Locality:} Spatially separated systems cannot instantaneously influence one another.
    \item[] \textbf{Completeness:} A complete theory must represent every element of physical reality.
\end{enumerate}
Given the quantum-mechanical correlations exhibited by entangled systems at spacelike separation, EPR concluded that at least one of these assumptions must fail if the formalism is to remain consistent. Einstein favored abandoning completeness, proposing instead that hidden variables determine the outcomes of measurements. The reasoning is as follows.

EPR considered a pair of particles prepared in an entangled state with perfectly correlated momenta and positions, such as
\[
\Psi(x_1, x_2) = \int dp\, e^{i(x_1 - x_2)p/\hbar}\,\phi(p),
\]
where the total momentum is conserved and $\phi(p)$ represents an arbitrary momentum-space wavefunction. Quantum mechanics allows one to predict the position $x_2$ of particle~2 with certainty by measuring the position $x_1$ of particle~1. By the reality criterion, $x_2$ therefore corresponds to an element of physical reality. By locality, the measurement performed on particle~1 cannot disturb particle~2, so this element of reality must have existed prior to measurement. Moreover, an analogous argument applies if one instead measures momentum.  However, quantum mechanics does not assign definite values to both position and momentum (nor any pair of non-commuting observables) simultaneously. EPR therefore concluded that the quantum-mechanical description does not represent all elements of physical reality and is thus incomplete.

\subsection{Bell’s Theorem}

\begin{figure*}
    \centering
    \includegraphics[clip, trim=5cm 6.2cm 5cm 11cm,width=\textwidth]{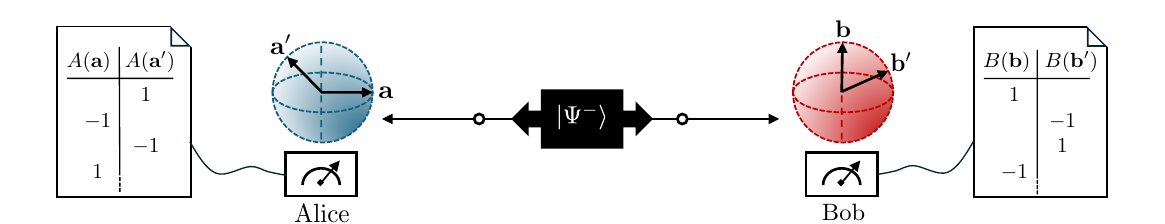}
    \caption{\label{fig:Bell}A schematic of the Bell--CHSH experiment. A source emits a spin-singlet pair to spacelike-separated stations. Alice chooses one of two pre-established measurement orientations, $\mathbf{a}$ and $\mathbf{a}'$, and respectively Bob with $\mathbf{b}$ and $\mathbf{b}'$. They record the orientation they chose as well as the outcome $A,B\in\{\pm1\}$ for each run. The observed statistics constructed from expectations $E(\mathbf{a},\mathbf{b})$ violate the CHSH bound (\ref{eq:CHSHineq}) for suitable settings.}
\end{figure*}

In 1964, John Bell provided the first precise and empirically testable formulation of the conflict between the EPR assumptions and the predictions of quantum mechanics \cite{Bell1964}. He showed that any theory supplementing quantum mechanics with local hidden variables must satisfy certain statistical constraints—\emph{Bell inequalities}—which are nevertheless violated by quantum predictions \cite{AspectExperimental1982}.

Consider two observers, Alice and Bob, performing measurements on a pair of entangled spin-\(\tfrac{1}{2}\) particles prepared in the singlet state
\begin{align}
\ket{\Psi^-} = \frac{1}{\sqrt{2}}\big(\ket{\uparrow}_A\ket{\downarrow}_B - \ket{\downarrow}_A\ket{\uparrow}_B\big). \nonumber
\end{align}
Each observer measures spin along a chosen direction, \(\mathbf{a}\) for Alice and \(\mathbf{b}\) for Bob, obtaining outcomes \(A,B \in \{-1,+1\}\). This setup is illustrated in Figure~\ref{fig:Bell}. In a local hidden-variable model, these outcomes are determined by a shared variable \(\lambda\), distributed according to a probability density \(p(\lambda)\), such that the expectation value factorizes:
\begin{align}
E(\mathbf{a},\mathbf{b}) = \int d\lambda\, p(\lambda)\, A(\mathbf{a},\lambda)\, B(\mathbf{b},\lambda). \nonumber
\end{align}

To derive Bell’s inequality in the form introduced by CHSH \cite{clauser1969proposed}, consider the combination
\begin{align}
S(\lambda) = A(\mathbf{a},\lambda)&           \big[B(\mathbf{b},\lambda) - B(\mathbf{b}',\lambda)\big] \nonumber \\
      & + A(\mathbf{a}',\lambda)\big[B(\mathbf{b},\lambda) + B(\mathbf{b}',\lambda)\big].\nonumber
\end{align}
Since by construction one bracket equals $\pm2$ and the other $0$, one finds \(|S(\lambda)| \leq 2\) for every \(\lambda\). Averaging over \(\lambda\) with a $p(\lambda)$ normalized to $1$ then yields the CHSH inequality \cite{clauser1969proposed},
\begin{align}\label{eq:CHSHineq}
|E(\mathbf{a},\mathbf{b}) - E(\mathbf{a},\mathbf{b'})| 
+ |E(\mathbf{a'},\mathbf{b}) + E(\mathbf{a'},\mathbf{b'})| \le 2.
\end{align}

Quantum mechanics predicts a different correlation. In the quantum description, $E(\mathbf{a},\mathbf{b})$ is computed as the expectation value of the corresponding spin operators in the joint quantum state. For the singlet state,\footnote{The final equality follows by expanding $(\boldsymbol{\sigma}_A\!\cdot\!\mathbf{a})(\boldsymbol{\sigma}_B\!\cdot\!\mathbf{b})=\sum_{i,j}a_i b_j\,\sigma_{A,i}\sigma_{B,j}$ and using the identity $\langle\Psi^-|\sigma_{A,i}\sigma_{B,j}|\Psi^-\rangle=-\delta_{ij}$ for the singlet state, which can be verified by direct calculation.}
\begin{align}
E(\mathbf{a},\mathbf{b})
&= \langle \Psi^- | \,(\boldsymbol{\sigma}_A \!\cdot\! \mathbf{a})\,(\boldsymbol{\sigma}_B \!\cdot\! \mathbf{b}) \,| \Psi^- \rangle \nonumber\\
&= -\,\mathbf{a}\!\cdot\!\mathbf{b} = -\cos\theta_{ab}, \nonumber
\end{align}
where $\boldsymbol{\sigma}_A$ and $\boldsymbol{\sigma}_B$ denote the Pauli operator vectors acting on Alice's and Bob's particles, respectively, and $\theta$ is the angle between the measurement directions. Choosing measurement settings such that the differences between directions\footnote{For example,  in a single plane: $\mathbf{a}=0^\circ$, $\mathbf{a'}=90^\circ$, $\mathbf{b}=45^\circ$, and $\mathbf{b'}=-45^\circ$ such that $\theta_{ab}=45^\circ$, $\theta_{ab'}=45^\circ$, $\theta_{a'b}=-45^\circ$, and $\theta_{a'b'}=135^\circ$.
} are \(45^\circ\) leads to a maximal quantum value of
\begin{align}
|E(\mathbf{a},\mathbf{b}) - E(\mathbf{a},\mathbf{b'})| + |E(\mathbf{a'},\mathbf{b}) + E(\mathbf{a'},\mathbf{b'})| = 2\sqrt{2} > 2, \nonumber
\end{align}
which violates (\ref{eq:CHSHineq}), thereby excluding any local hidden variable explanations. Such maximal quantum violations of Bell inequalities are known as Tsirelson bounds \cite{Tsirelson1980}. In fact, the inequality is violated for a continuous range of relative angles, not only at this maximal setting. 

Numerous experiments—culminating in so-called \emph{loophole-free} tests \cite{Hensen2015,LyndenLoophole2015,GiustinaLoophole2015}—have confirmed these violations. Such experiments are designed to exclude all known classical explanations by ensuring spacelike separation of measurements, high detection efficiency, and independent choice of measurement settings. Bell’s theorem therefore establishes that no theory maintaining both locality and realism can reproduce the empirical predictions of quantum mechanics.

\subsection{Hardy’s Paradox}\label{subsec:Hardy}

Lucien Hardy proposed a striking variation of Bell’s argument that reveals quantum nonlocality without relying on inequalities \cite{HardyNonlocality1993}. Hardy considered an entangled two-particle state arranged so that certain joint outcomes are strictly forbidden in classical local models yet occur with nonzero probability in quantum mechanics.

Let two observers choose between two dichotomic measurements, \(A_1, A_2\) for Alice and \(B_1, B_2\) for Bob, with outcomes \(\pm1\). In a local-realist model, the values \(A_i(\lambda)\) and \(B_j(\lambda)\) are predetermined. Hardy showed that for a suitably chosen entangled state (one example being),
\[
\ket{\psi_H} = \frac{1}{\sqrt{3}}\big(\ket{\uparrow}_A\ket{\downarrow}_B + \ket{\downarrow}_A\ket{\uparrow}_B + \ket{\downarrow}_A\ket{\downarrow}_B\big),
\]
quantum mechanics predicts a non-zero probability \(P(A_1=+1,B_1=+1)>0\), even though the logical constraints implied by local realism would make this outcome impossible.

Specifically, one can choose measurement settings such that
\begin{align}
&(a) \quad P(A_2=+1,B_2=+1)=0, \nonumber \\ \quad
&(b)  \quad P(A_1=+1,B_2=-1)=0, \nonumber\\
&(c) \quad P(A_2=-1,B_1=+1)=0. \nonumber
\end{align}
Under the assumption of predetermined local values, the conditions $(a)$--$(c)$ together imply\footnote{Let us assert that $A_1=+1$. From $(b)$, if $A_1=+1$, then $B_2 = +1$. From $(a)$, if  $B_2 = +1$, then $A_2=-1$. Lastly, from $(c)$, $A_2=-1$ entails $B_1=-1$, prohibiting the possibility of $B_1=+1$. The reverse logical order (starting with $B_1=+1$) concludes similarly}
\[
(d) \quad P(A_1=+1,B_1=+1)=0,
\]
since any assignment of $A_1=+1$ or $B_1=+1$ contradicts the other. However, quantum mechanics predicts\footnote{This can be shown using the example measurement settings: setting $2$ along the $\{ \ket{\uparrow},\ket{\downarrow}\}$ axis, recording the spin $\ket{\uparrow}$ as $+1$ and $\ket{\downarrow}$ as $-1$ (so that outcome $A_2=+1$ corresponds to state $\ket{\uparrow}$); setting $1$ measuring an orthogonal spin axis, recording $\ket{+}=(\ket{\uparrow}- \ket{\downarrow})/\sqrt{2}$ 
as $+1$ and $\ket{-}=(\ket{\uparrow}+ \ket{\downarrow})/\sqrt{2}$ 
as $-1$ (so that outcome $A_1=-1$ corresponds to $\ket{-}$). The probability pairings are then given by the square of the relevant overlap, e.g., $P(A_1=+1,B_2=-1)=\braket{+,\downarrow}{\psi_H}^2$.
}
\[
(d^\dagger) \quad P(A_1=+1,B_1=+1)=\tfrac{1}{12}.
\]
An event that cannot occur in a local realist framework is nevertheless allowed by quantum theory\footnote{The maximum two-qubit Hardy probability is $(5\sqrt{5}-11)/2$, achieved by a non-symmetric state and settings \cite{HardyNonlocality1993}.}. Hardy’s paradox thus provides a particularly transparent demonstration of nonlocality, 
arising from logical inconsistency rather than statistical inequalities.

\subsection{The de Broglie--Bohm Theory}

An alternative response to the EPR dilemma is to retain realism and determinism by abandoning locality. The de Broglie–Bohm theory \cite{1927deBroglie,1Bohm1952,2Bohm1952}, or pilot-wave theory, adopts this strategy and resolves the wave–particle duality puzzle by positing both: a real wavefunction guiding particles that always possess definite positions \cite{BellImpossible1982,Holland_1993,goldstein1998quantum2}.

In this framework, the wavefunction \(\psi(\mathbf{r}_1,\dots,\mathbf{r}_N,t)\) evolves according to the Schrödinger equation as usual, but each particle follows a deterministic trajectory \(\mathbf{r}_k(t)\) satisfying the guiding equation \cite{1Bohm1952}
\[
\frac{d\mathbf{r}_k}{dt} = \frac{\hbar}{m_k}\,\Im \frac{\nabla_k \psi}{\psi}(\mathbf{r}_1,\dots,\mathbf{r}_N,t),
\]
where \(\Im\) denotes the imaginary part. The velocity of each particle depends on the instantaneous configuration of all others through the wavefunction, implying explicit nonlocality.


Statistical predictions coincide with those of standard quantum mechanics provided that the distribution \(p(\mathbf{r}_1,\dots,\mathbf{r}_N,t)\) of particle configurations satisfies the \emph{quantum equilibrium} condition
\[
p(\mathbf{r}_1,\dots,\mathbf{r}_N,t) = |\psi(\mathbf{r}_1,\dots,\mathbf{r}_N,t)|^2,
\]
which is preserved by the dynamics once imposed as an initial condition. Under quantum equilibrium, the theory exactly reproduces the Born rule, though it allows, in principle, for deviations if non-equilibrium distributions could be realized.

The de Broglie–Bohm theory restores determinism and a clear particle ontology, providing a straightforward resolution of the measurement problem in which systems always possess definite positions and outcomes arise from the dynamics rather than postulated collapse. These explanatory gains come at a cost: the guiding dynamics is explicitly nonlocal and difficult to reconcile with relativistic spacetime structure \cite{BerndlNonlocality1996}, although various solutions have been proposed \cite{DurrRelativisticBohmian2014,DurrHypersurface1999,Bell1987Beables,Sutherland2022}; particle positions play a privileged role as the only variables that possess definite values independent of measurement; 
and the quantum equilibrium condition is typically assumed rather than derived from deeper principles. As a result, the theory offers a coherent but 
ontologically richer realist alternative to orthodox quantum mechanics.

\section{The Measurement Problem and Contextuality}\label{sec:MeasurementProblem}

The results of the preceding discussion place severe constraints on any attempt to understand quantum mechanics in classical terms. The EPR argument, Bell’s theorem, and related results show that no theory maintaining both locality and realism can reproduce the empirical predictions of quantum mechanics, while approaches such as de Broglie–Bohm theory illustrate that restoring deterministic realism requires accepting fundamental nonlocality. In light of these developments, it is natural to shift the interpretational focus from questions of completeness and locality to a more basic problem: how definite measurement outcomes arise at all from a linear, deterministic formalism. This lies at the heart of the measurement problem. Closely related is the phenomenon of contextuality, which shows that measurement outcomes cannot be understood as revealing pre-existing properties of a system independent of the experimental context. Together, these issues probe the relationship between the quantum formalism and the emergence of classical reality.

\subsection{Contextuality}

The notion of \emph{contextuality} addresses whether the value of a physical observable can be assigned independently of the measurement context in which it is embedded \cite{Kok2023}.
A noncontextual hidden-variable theory assumes that the value of each observable \(A\) is predetermined, and that measuring \(A\) as part of different compatible sets of observables\footnote{Here, a \emph{measurement context} refers to a set of mutually compatible (i.e.\ commuting) observables that can be jointly measured. Importantly, the same observable may belong to more than one such compatible set. For example, in a spin-1 system, the observable \(S_z^2\) commutes with both \(S_x^2\) and \(S_y^2\), even though \(S_x^2\) and \(S_y^2\) do not commute with each other.} should always yield the same result.

The Kochen–Specker theorem (KS) \cite{KochenSpecker1967} demonstrates that noncontextual value assignments of this kind are impossible for quantum systems of dimension three or higher. The proof proceeds by contradiction: assuming that every observable possesses a context-independent value that respects the functional relations among commuting observables leads to an inconsistency with the structure of Hilbert space. Rather than reproducing the original KS construction, we illustrate this result using a particularly transparent state-independent argument due to Peres and Mermin \cite{PERES1990107,MerminSimple1990,MerminHidden1993}.



Consider the following set of nine observables acting on a two-qubit system, arranged in a $3\times3$ array:
\[
\begin{array}{ccc}
\sigma_x\otimes \mathbb{I} & \mathbb{I}\otimes\sigma_x & \sigma_x\otimes\sigma_x\\
\mathbb{I}\otimes\sigma_y & \sigma_y\otimes \mathbb{I} & \sigma_y\otimes\sigma_y\\
\sigma_x\otimes\sigma_y & \sigma_y\otimes\sigma_x & \sigma_z\otimes\sigma_z
\end{array}
\]
Each row and each column consists of mutually commuting observables and therefore defines a valid measurement context. Under noncontextual assumptions, each observable $A$ is assigned a predetermined value $v(A)\in\{\pm1\}$, independent of context, and these assignments respect functional consistency: for any commuting observables $A$ and $B$, one has $v(AB)=v(A)v(B)$.

Quantum mechanically, the product of the observables in each row is $+\mathbb{I}$, and the product of the observables in the first two columns is also $+\mathbb{I}$, but the product in the third column is $-\mathbb{I}$. Functional consistency therefore requires that the product of the assigned values in each row be $+1$, and likewise for the first two columns, while the product in the final column must be $-1$.

Any noncontextual value assignment \(v(A)\) would require that the product of all rows and columns yield \(+1\) since each observable appears exactly twice and $v(A)^2=1$, yet quantum mechanics predicts \(-1\). This contradiction shows that no noncontextual value assignment is possible. The conclusion is that quantum mechanics is inherently contextual.

\subsection{Contextuality and Bell's Theorem }

To further elucidate on quantum contextuality and its relation to Bell nonlocality we provide a state-dependent Hardy-like paradox from Cabello et al. \cite{CabelloHardy2013} in which quantum mechanics predicts a non-zero probability for an outcome that is forbidden by noncontextual realist accounts.

Consider a single physical system about which we can make five claims, $A_1,...,A_5$, that are either true or false ($T$/$F$). Suppose the system is prepared such that the following joint probabilities vanish:
\begin{align}
    (a^*) & \quad P(A_1=T,A_2=T)=0 \nonumber \\
    (b^*) & \quad P(A_2=F,A_3=F)=0 \nonumber \\
    (c^*) & \quad P(A_3=T,A_4=T)=0 \nonumber \\
    (d^*) & \quad P(A_4=F,A_5=F)=0 . \nonumber
\end{align}
In a noncontextual realist model, each $A_i$ is assumed to possess a predetermined truth value, independent of which other compatible statements are jointly considered. Under this assumption, the constraints $(a^*)$--$(d^*)$ imply
\begin{align}
    (e^*) \quad P(A_1=T,A_5=F)=0 \nonumber ,
\end{align}
since any assignment in which $A_1$ is true necessarily forces $A_5$ to be true as well.\footnote{The reasoning is as follows: by $(a^*)$, if $A_1$ is true then $A_2$ must be false; by $(b^*)$, if $A_2$ is false then $A_3$ is true; by $(c^*)$, if $A_3$ is true then $A_4$ is false; and by $(d^*)$, if $A_4$ is false then $A_5$ must be true.}

Quantum mechanically, let us identify $\ket{A_1}$ as the state of a system with property $A_1 = T$. We then represent each proposition $A_i$ by a projector $\Pi_i=\ket{A_i}\!\!\bra{A_i}$. The probability that $A_i$ is true in a state $\ket{\psi}$ is $P(A_i=T)=|\!\! \braket{A_i}{\psi}\!\!|^2 =\bra{\psi}\Pi_i\ket{\psi}$, while $P(A_i=F)=\bra{\psi}(\mathbb{I}-\Pi_i)\ket{\psi}$. Joint probabilities such as $P(A_i=T,A_j=T)$ are given by $\bra{\psi}\Pi_i \Pi_j\ket{\psi}$, where the relevant projectors commute, reflecting the compatibility of the corresponding measurements.

For the three-level quantum system prepared in the state
\[
\ket{\psi}=\frac{1}{\sqrt{3}}\begin{pmatrix}1\\1\\1\end{pmatrix},
\]
one can choose projectors $\Pi_1,\dots,\Pi_5$ satisfying the commutation relations associated with $(a^*)$--$(d^*)$\footnote{For example, take $A_1=\frac{1}{\sqrt{3}}(1,-1,1)^T$, $A_2=\frac{1}{\sqrt{2}}(1,1,0)^T$, $A_3=(0,0,1)^T$, $A_4=(1,0,0)^T$, and $A_5=\frac{1}{\sqrt{2}}(0,1,1)^T$.} such that quantum mechanics predicts
\begin{align}
    (e^\dagger) \quad P(A_1=T,A_5=F)=\frac{1}{9}>0 \nonumber ,
\end{align}
even though this outcome is forbidden by the noncontextual realist inference above. Thus, an event that cannot occur in a noncontextual realist framework is nevertheless allowed by quantum theory. This contradiction demonstrates that quantum mechanics is contextual.

By comparing the above argument with Section~\ref{subsec:Hardy}, contextuality is seen to be closely related to the failure of classical realism revealed by Bell’s theorem. Whereas Bell inequalities probe the compatibility of quantum correlations with locality and noncontextual value assignments, contextuality addresses more generally whether measurement outcomes can be understood as revealing pre-existing, measurement-independent properties. From this perspective, Bell nonlocality can be viewed not as a dynamical faster-than-light influence, but as a particular form of contextuality arising in spacelike-separated measurement contexts.
Taken together, they rule out a broad class of realist hidden-variable theories, leaving only explicitly nonlocal and
contextual alternatives.



\subsection{The Measurement Problem}\label{subsec:measurementProblem}

The measurement problem concerns the apparent conflict between the continuous, deterministic evolution of the wavefunction under the Schrödinger equation and the discontinuous, probabilistic nature of measurement outcomes \cite{vNmeasurement,schlosshauer2007decoherence,Maudlin1995}. Formally, the postulates of quantum mechanics provide two distinct rules for the evolution of a system:

\begin{itemize}
    \item[]\textbf{Postulate 2:} Continuous unitary evolution according to (\ref{eq:TDSE}) which preserves superpositions.
    \item[]\textbf{Postulate 5:} Discontinuous state reduction, in which measurement of an observable \(S\) with eigenstates \(\{\ket{s_i}\}\) and outcome \(s_j\) produces $\ket{\psi} \longrightarrow \ket{s_j}$ with probability $P(s_j) = |\langle s_j|\psi\rangle|^2$ in line with the Born rule (\ref{eq:BornRule}).
\end{itemize}

The problem is that the Schrödinger dynamics, when applied to the joint system of the object and measuring apparatus, predicts that the two become entangled rather than yielding a definite outcome. In von Neumann’s measurement scheme \cite{vNmeasurement}, the interaction is represented as
\begin{align}\label{eq:VNMeasurementInteraction}
\sum_i c_i \ket{s_i}\ket{\Phi_0} \longrightarrow \sum_i c_i \ket{s_i}\ket{\Phi_i},
\end{align}
where \(\ket{\Phi_0}\) is the apparatus initial ``ready'' state and each \(\ket{\Phi_i}\) is a distinct \emph{pointer state} of the apparatus corresponding to the system state $\ket{s_i}$. The final state is a superposition of macroscopically distinct outcomes, yet in experience we only ever observe one. This is the essence of the measurement problem. Schrödinger illustrated this tension with his famous cat thought experiment, in which a microscopic superposition is amplified to the macroscopic scale, seemingly implying an absurdity: a cat that is simultaneously alive and dead until observed \cite{SchrodingerCat1935}. The challenge is to explain how and when a single outcome is selected.


{It is useful to decompose the measurement problem into related but distinct issues \cite{schlosshauer2007decoherence}. These issues may be constructed from four observations about measurement:}

\begin{enumerate}
    \item[(i)] \textit{Preferred bases} — 
    The unitary dynamics of quantum mechanics does not single out a unique decomposition of an entangled state into system–apparatus correlations. For example, the state in (\ref{eq:VNMeasurementInteraction}) can equally be expressed in another orthonormal basis as $\sum_j d_j \ket{s'_j}\ket{\Phi'_j}$. Why, then, do measurements yield definite outcomes in a particular quasi-classical basis (such as position), rather than in arbitrary superpositions of pointer states?

    \item[(ii)] \textit{Coherence suppression} — Although unitary evolution preserves phase relations between branches of the wavefunction, interference between macroscopically distinct states is not observed under ordinary conditions. Why are such coherence phenomena effectively absent at macroscopic scales?

    \item[(iii)] \textit{Agreement and objectivity} — Why do different observers, independently accessing the same system, reliably agree on the outcome obtained?

    \item[(iv)] \textit{Definite outcomes} — Why does a single outcome occur in each measurement, rather than the post-measurement state remaining a superposition of alternative outcomes weighted by the Born rule? And, if only one outcome is realized, by what physical mechanism is it selected?

\end{enumerate}

Several answers have been proposed in response to these issues. One is to interpret the wavefunction as representing knowledge rather than reality, as in the epistemic reading of the Copenhagen interpretation. Another is to modify the dynamics so that collapse occurs objectively, independent of observation. A third is to seek an explanation of apparent collapse through environmental decoherence, which we will examine, alongside other influential theories, in Section~\ref{sec:decoherence}. {Each response can be benchmarked by how well it addresses (i)–(iv).}

\subsection{Objective Collapse Models}

Objective collapse models \cite{Gao_2018,collapseBassi,BASSI2003257,BassiCollapseTheoryExperiment2013} modify the Schrödinger equation to include spontaneous, stochastic collapses of the wavefunction. The most well-known example is the Ghirardi--Rimini--Weber (GRW) model \cite{GhirardiGRW1986}. In GRW, each particle undergoes spontaneous localization in position space at random times, with a mean frequency \(\lambda \approx 10^{-16}\,\mathrm{s}^{-1}\). Between collapses, the wavefunction evolves unitarily.

The GRW dynamics for an \(N\)-particle wavefunction \(\psi(\mathbf{r}_1,\dots,\mathbf{r}_N,t)\) can be written schematically as \cite{GhirardiGRW1986}
\[
d\psi = -\frac{i}{\hbar}H\psi\,dt + \sum_{k=1}^N \left( \frac{L_k (\mathbf{x}_k) }{\|L_k (\mathbf{x}_k)\psi\|} - \mathbb{I}\right)\psi\, dN_k(t),
\]
where \(L_k (\mathbf{x}_k) \) is a localization operator acting on the \(k\)th particle, and \(dN_k(t)\) is a {Poisson counting-process governing random collapse. Thus, each increment \(dN_k(t)\) is $0$ if there is no hit of type $k$ in the time window, and $1$ if there is a hit. When a collapse occurs ($dN_k(t) = 1$), the wavefunction is multiplied by a narrow Gaussian centered at a random position \(\mathbf{x}_k\), $\psi \rightarrow L_k(\mathbf{x}_k) \psi / \norm*{L_k(\mathbf{x}_k) \psi}$,} ensuring that macroscopic superpositions are rapidly suppressed.

An alternative, continuous version is the CSL (Continuous Spontaneous Localization) model, in which collapses occur gradually rather than instantaneously \cite{GhirardiCSL1990}:
\begin{align}
d\psi = \bigg[-\frac{i}{\hbar}H\,dt + \sqrt{\lambda}\int d^3x\,[M(\mathbf{x})-\langle M(\mathbf{x})\rangle]\, dW(\mathbf{x},t) \nonumber\\
- \frac{\lambda}{2}\int d^3x\,[M(\mathbf{x})-\langle M(\mathbf{x})\rangle]^2 dt \bigg]\psi, \nonumber
\end{align}
where \(M(\mathbf{x})\) is the mass-density operator and \(dW(\mathbf{x},t)\) is a Wiener process representing stochastic noise. 

{Objective collapse models provide a genuine dynamical resolution of the measurement problem: superpositions of macroscopically distinct states are rendered physically unstable, and definite outcomes arise from modified dynamics rather than interpretational stipulation. In doing so, they address issues (i), (ii) and (iv)
within a single framework. Crucially, these models are empirically distinguishable from standard quantum mechanics. While deviations are negligible for microscopic and macroscopic systems, small but in principle observable effects are predicted in the mesoscopic regime, where collapse rates become appreciable while coherence remains experimentally accessible \cite{BassiCollapseTheoryExperiment2013}. This testability has already been exploited to place meaningful constraints on the parameter space of the CSL model \cite{Bassam2017,Vinante2017}, and thus constitutes a significant strength of the approach. The price, however, is the introduction of new fundamental constants and stochastic terms not derived from deeper principles, as well as ongoing challenges in constructing fully relativistic and quantum field theoretic formulations \cite{Jones_2021}. Objective collapse theories thus trade theoretical economy for dynamical clarity and empirical falsifiability.}

\section{Decoherence \& the Emergence of Classicality}\label{sec:decoherence}

The preceding discussion highlighted two central challenges in quantum foundations: the contextual nature of quantum observables and 
the tension between the deterministic evolution of the wavefunction and the appearance of definite  measurement outcomes. While objective collapse models address this tension by modifying the dynamics of quantum theory, an alternative approach retains the standard formalism and instead examines the role of environmental interactions in shaping observed phenomena. This approach is known as \emph{decoherence}~\cite{Joos1985,SchlosshauerInterp2005}.

Decoherence provides a physical mechanism by which interference between distinct quantum alternatives is dynamically suppressed through entanglement with environmental degrees of freedom, giving rise to the appearance of classical behavior. Building on this insight, this section examines two interpretational frameworks that take decoherence as central: the \emph{many-worlds interpretation}, which treats the universal wavefunction as real and branching, and the \emph{consistent-histories framework}, which generalizes quantum mechanics to sequences of events without invoking wavefunction collapse.





\subsection{Density Operators and Open Quantum Systems} \label{subsec:densityOperators}

{Before discussing decoherence explicitly, it is useful to introduce a more general description of quantum states than that provided by state vectors alone\cite{vonNeumann1927,vonNeumann1932,Nielsen2000,Breuer2002}}. The standard formulation of quantum mechanics often considers isolated systems that are completely specified by a normalized state vector $|\psi\rangle$. Such states are known as \emph{pure states}. For a pure state, all statistical predictions may equivalently be expressed using the projection operator
\begin{equation}
\rho = |\psi\rangle\!\langle\psi|,
\end{equation}
known as the \emph{density operator} (or density matrix). Expectation values of observables may then be written in the compact form
\begin{equation}
\langle A\rangle
= \langle\psi|A|\psi\rangle
= \mathrm{Tr}(A\rho),
\end{equation}
which provides a formulation independent of any particular basis.

{The density operator becomes essential when our knowledge of a system is incomplete. A system may be prepared in different quantum states with classical probabilities, or may interact with external degrees of freedom that are not experimentally accessible. In such situations the system is described not by a single state vector but by a \emph{statistical mixture}~\cite{vonNeumann1927,vonNeumann1932},
\begin{equation}
\rho = \sum_i p_i |\psi_i\rangle\!\langle\psi_i|,
\qquad
\sum_i p_i = 1,
\end{equation}
where, for probabilistic preparation, $p_i$ denotes the probability that the preparation procedure produces the state $|\psi_i\rangle$. The coefficients $p_i$ therefore represent classical probabilistic uncertainty about which state has been prepared, rather than quantum superposition. The diagonal elements of $\rho$ represent populations, while off-diagonal elements encode phase coherence between different components of a superposition and therefore determine the system's ability to exhibit interference.}

{More generally, such mixtures may arise either from incomplete classical knowledge of the preparation (\emph{proper mixtures}) or from entanglement with degrees of freedom that are not observed (\emph{improper mixtures}). Although these situations differ conceptually, they are described operationally by the same density-operator formalism.}

{This formalism therefore provides a unified description of both pure and mixed states. Pure states satisfy $\rho^2=\rho$, whereas mixed states obey $\rho^2\neq\rho$. Time evolution of a closed system is governed by the von Neumann equation,
\begin{equation}
\frac{d\rho}{dt}
=
-\frac{i}{\hbar}[H,\rho],
\end{equation}
which is equivalent to Schr\"odinger evolution expressed in operator form.}

{In realistic situations, however, physical systems are rarely isolated. Instead they interact with surrounding environments whose microscopic degrees of freedom cannot be fully monitored. The combined system--environment state may still evolve unitarily, but observations typically access only the subsystem of interest. The appropriate description is therefore obtained by averaging over environmental degrees of freedom, a procedure known as the \emph{partial trace}:
\begin{equation}
\rho_S = \mathrm{Tr}_E(\rho_{SE}).
\end{equation}
The resulting reduced density operator $\rho_S$ describes an \emph{open quantum system}~\cite{Breuer2002}. Even when the total system--environment state remains pure, the reduced state of the subsystem generally becomes mixed. As we now show, this loss of coherence arising from entanglement with unobserved degrees of freedom provides the physical mechanism underlying decoherence.}

\subsection{Decoherence}

{Using the density-operator formalism introduced above, we now consider how a system interacting with an environment undergoes decoherence \cite{SchlosshauerInterp2005,DecoherenceRough}.} In standard quantum mechanics, a system and its environment become entangled through interaction. If we trace over the environmental degrees of freedom, the reduced density matrix of the system does not correspond to a pure superposition but rather a mixed state. Let the composite state of system \(S\) and environment \(E\) be
\[
\ket{\Psi_{SE}} = \sum_i c_i \ket{s_i}\ket{e_i},
\]
where the environmental states \(\ket{e_i}\) are correlated with distinct system states \(\ket{s_i}\). The reduced state of the system is then
\[
\rho_S = \mathrm{Tr}_E\big(\!\ket{\Psi_{SE}}\!\!\bra{\Psi_{SE}}\!\big)
      = \sum_{i,j} c_i c_j^* \ket{s_i}\!\!\bra{s_j} \braket{e_j}{e_i}.
\]
{Since orthogonal system states leave distinguishable and effectively irreversible records in many environmental subsystems}, the overlaps $\braket{e_j}{e_i}$ rapidly decay and environmental states become effectively orthogonal, \(\braket{e_i}{e_j} \approx \delta_{ij}\).\footnote{{This behavior can be seen by considering explicit system-environment coupling models \cite{ZurekEinselection1982,SpinEnvironmentsCucchietti2005}.}} When this happens, the off-diagonal interference terms vanish:
\[
\rho_S \approx \sum_i |c_i|^2 \ket{s_i}\!\!\bra{s_i}.
\]
This is the essence of decoherence: interference between macroscopically distinct states is dynamically suppressed through entanglement with the environment. The system therefore appears to have collapsed into a statistical mixture, even though the total system–environment state remains a pure superposition. 
{Importantly, this reduced state represents an \emph{improper} mixture arising from entanglement with the environment rather than classical ignorance about which outcome has actually occurred.}

Decoherence therefore provides a physical mechanism addressing two central aspects of the measurement problem \cite{Zurekreview2003,SchlosshauerInterp2005}. First, it provides an explanation for (i) the emergence of \emph{preferred bases}: environmental interactions dynamically single out pointer states---defined through the lens of decoherence theory as system states that are minimally disturbed by environmental interactions and therefore remain robust under decoherence---a process known as environment-induced superselection (einselection) \cite{ZurekEinselection1982}. Second, it accounts for (ii) \emph{coherence suppression} by showing why interference between macroscopically distinct states is effectively unobservable under ordinary conditions.

Building on this insight, the program of \emph{quantum Darwinism} \cite{ZurekqDarwinism} extends decoherence to address (iii) \emph{agreement and objectivity}. Information about pointer states is redundantly encoded across many independent fragments of the environment, allowing multiple observers to access the same classical information without significantly disturbing the system.

However, decoherence alone does not resolve (iv), the problem of \emph{definite outcomes}. Although interference becomes negligible and branches evolve independently, the total quantum state still contains a superposition of all alternatives. Decoherence explains the dynamical emergence of states that behave operationally like classical mixtures, 
but it does not by itself explain why only one outcome is realized in a given experimental run. It therefore provides a partial resolution of the measurement problem, addressing issues (i)–(iii) while leaving (iv) outstanding. As a result, many interpretational programs adopt decoherence as a foundational component, with a principal task being to supply a satisfactory account of definite outcomes.


\subsection{The Many-Worlds Interpretation}



The \emph{many-worlds interpretation} \cite{EverettMWI1957} accepts the universal wavefunction as a complete description of reality and denies that any physical collapse occurs. The Schrödinger equation (\ref{eq:TDSE}) is assumed to be universally valid, applying without exception to microscopic systems, measuring devices, and observers alike. Apparent collapse is reinterpreted as a consequence of entanglement and subsequent decoherence.

As per (\ref{eq:VNMeasurementInteraction}), when a measurement interaction produces an entangled superposition of system–apparatus states, \(\ket{s_i}\ket{\Phi_i}\), decoherence renders the resulting branches dynamically autonomous. Each branch corresponds to a distinct quasi-classical history in which a definite outcome is recorded. {From this perspective, the reduced density matrices produced by decoherence are interpreted not as classical ignorance mixtures but as arising from the observer’s entanglement with dynamically autonomous branches of the universal wavefunction.} 
All branches persist, but observers within a given branch experience a single definite result. In this way, many-worlds accepts the full superposition predicted by unitary dynamics yet negates issue (iv) by denying the fundamental selection of a unique outcome.

The many-worlds interpretation offers a dynamically simple account of quantum phenomena by retaining universal unitarity and dispensing with physical collapse, but it faces a number of significant conceptual challenges. Notable among these is the interpretation of probability in a deterministic (branching) multiverse in which all outcomes occur \cite{KentAgainstMWI,GreavesProb2007,Albert2010}. In particular, it is unclear how the Born rule is to be justified: frequentist notions appear ill-suited to a setting where branches do not form a countable ensemble accessible to observers, while alternative approaches must explain why branch weights should be given by \(|c_i|^2\) rather than some other measure. One influential proposal, developed by Deutsch and Wallace \cite{DeutschDecisions1999,WALLACE2007311,Wallace2012book}, appeals to decision theory, arguing that a rational agent embedded in a branching universe should assign subjective probabilities proportional to \(|c_i|^2\). Subject to ongoing debate, other approaches appeal to symmetry considerations \cite{ZurekManyWorlds?} or to forms of self-locating uncertainty \cite{SaundersBranching2008,SebensSelflocating2018}.

Beyond the probability problem, further concerns include the approximate and emergent nature of branching in decoherence-based models. While decoherence suppresses interference on extremely short timescales, it does so only approximately, leaving no sharp criterion for when branching occurs or how distinct branches are to be precisely defined. Related issues concern the status of observers and personal identity across branching \cite{SaundersBranching2008,GREAVES2004}, as well as the ontological cost of postulating a vast multiplicity of equally real branches. While the many-worlds interpretation provides a coherent and mathematically conservative framework, its conceptual foundations continue to be the subject of active debate.

\subsection{Consistent Histories}

The \emph{consistent-histories} or \emph{decoherent-histories} approach \cite{GriffithsConsistent1984,OmnesLogical1988,GellMannHartleClassical1993} generalizes quantum mechanics by assigning probabilities to entire sequences of events—called \emph{histories}—without making fundamental reference to measurements or external observers. 

A history is represented by a time-ordered sequence of projection operators
\[
C_\alpha = P_{\alpha_n}(t_n)\dots P_{\alpha_1}(t_1),
\]
acting on the initial state \(\rho_0\). The decoherence functional
\[
D(\alpha,\beta) = \mathrm{Tr}\!\left(C_\alpha \rho_0 C_\beta^\dagger\right)
\]
measures the interference between two histories \(\alpha\) and \(\beta\).  
A set of histories is said to be \emph{consistent} (or decoherent) if the interference terms vanish for distinct histories,
\[
D(\alpha,\beta) = 0 \quad \text{for} \quad \alpha \neq \beta.
\]
Such a set—often called a \emph{framework} \cite{GriffithsConsistent1984}—is a collection of mutually exclusive and exhaustive histories to which classical probability rules may be consistently applied.\footnote{{For example, consider a single qubit prepared in $\ket{0}$. At a later time $t$, one framework is defined by the commuting projectors 
$\{ \ket{0}\!\bra{0},\,\ket{1}\!\bra{1} \}$, yielding exclusive histories corresponding to definite $z$-spin. An incompatible alternative framework is defined by $\{ \ket{+}\!\bra{+},\,\ket{-}\!\bra{-} \}$, where $\ket{+}=(\ket{0}+\ket{1})/\sqrt{2}$. Each framework, taken by itself, permits consistent probability assignments.}} Within a given framework, exactly one history occurs and probabilities may be assigned according to
\[
p(\alpha) = D(\alpha,\alpha)
\]
as the probability of history \(\alpha\). The consistency condition is typically justified physically through decoherence mechanisms, which suppress interference between alternative histories and thereby permit classical probability assignments.

The consistent-histories approach encompasses both standard quantum mechanics and classical stochastic theories as limiting cases and is particularly well suited to closed systems, such as the early universe \cite{GellMannCosmo1990}, where the notion of an external observer is inappropriate. Like the many-worlds interpretation, the formalism dispenses with wavefunction collapse, but unlike Everett’s picture it treats histories within a given consistent framework as mutually exclusive alternatives rather than coexisting realities. In this way, issue (iv) is addressed by stipulating that, within any single framework, exactly one history occurs. 
This resolution relies crucially on the \emph{single-framework rule}, which prohibits combining incompatible descriptions and thereby also avoids contradictions associated with contextuality and Bell-type correlations \cite{goldstein1998quantum}. The cost of this strategy is that the formalism permits multiple incompatible but equally valid frameworks, so no unique, framework-independent account of past quantum events is selected. Whether this reflects a faithful representation of quantum structure or an unsatisfactory restriction on physical explanation remains contested.

\section{Conclusion}
This introduction, following the structure depicted in Figure~\ref{fig:decision-tree}, has examined some of the ways in which the formalism of quantum mechanics constrains the range of viable interpretations. Beginning with the standard postulates and the Copenhagen tradition, we saw that the operational success of quantum mechanics leaves open fundamental questions about the status of the quantum state, the role of probability, and the nature of measurement. These questions motivate the distinction between epistemic and ontic conceptions of the wavefunction, a distinction sharpened by the recent no-go result of PBR. The PBR theorem places strong constraints on $\psi$-epistemic hidden-variable models, demonstrating that, under preparation independence, distinct quantum states cannot correspond to the same underlying physical reality.

The EPR argument and Bell’s theorem showed that any realist completion of quantum mechanics must abandon locality, while Hardy’s paradox further clarified the logical tension between local realism and quantum predictions. The de Broglie–Bohm theory illustrates that determinism and realism can be retained only by embracing explicit nonlocality.

The discussion of contextuality and the measurement problem highlighted deeper limitations on classical intuitions about physical properties and dynamical evolution. Contextuality shows that measurement outcomes cannot be understood as revealing pre-existing, measurement-independent values, while the measurement problem exposes the tension between unitary dynamics and definite outcomes. Objective collapse models address this tension by modifying the dynamics with stochastic influences, whereas decoherence provides a physical mechanism for the suppression of interference and the emergence of classical behavior without introducing collapse.

Finally, interpretations such as many-worlds and consistent histories demonstrate how decoherence can be incorporated into broader conceptual frameworks that retain unitary evolution, though at the cost of significant interpretational commitments concerning probability, ontology, or the status of alternative histories. Across all approaches, the central lesson is that the empirical success of quantum mechanics comes with unavoidable trade-offs between locality, realism, determinism, and explanatory completeness.

Other approaches and foundational results, not discussed in this introductory material, further enrich the interpretational landscape. Alternative realist programs include Nelson’s stochastic mechanics \cite{Nelson1966} and superdeterministic approaches \cite{tHooft2007, Palmer2020}, the latter of which evade Bell-type constraints by rejecting statistical independence. Other perspectives, such as relational quantum mechanics \cite{Rovelli1996}, 
hold that quantum states and physical properties are only meaningful relative to other physical systems, rather than as absolute observer-independent quantities. Additional no-go theorems---including the no-cloning \cite{Wootters1982}, no-broadcasting \cite{Barnum1996}, and no-deleting theorems \cite{Pati2000}---likewise continue to highlight ways in which quantum mechanics departs fundamentally from classical intuitions about information, measurement, and physical reality.

Rather than singling out a uniquely correct interpretation, the results surveyed here delineate the structure of the problem space itself. Quantum foundations thus remains a field in which progress is driven not only by new experiments \cite{Carlesso2022}, but also by clarifying which combinations of assumptions are jointly untenable. In this sense, foundational analysis continues to play a central role in understanding what quantum mechanics does—and does not—tell us about physical reality.

\section*{Acknowledgments}
TM is supported by the Engineering and Physical Sciences Research Council [grant number EP/W524347/1]. AN acknowledges support and hospitality from the organisers of the 33rd Chris Engelbrecht Summer School, Stellenbosch 2025.

\bibliographystyle{ieeetr} 
\bibliography{refs.bib}
\end{document}